\documentclass[prl,reprint,amsmath,amssymb,aps]{revtex4-1}
     \usepackage{color}
\usepackage{graphicx}
\usepackage{dcolumn}
\usepackage{bbm}
\usepackage{subcaption}

\renewcommand{\geq}{\geqslant}

\newcommand{\ms}{m_{\rm s}}
\newcommand{\mus}{\mu_{\rm s}}

\newcommand{\E}{\mathbf{E}}
\newcommand{\K}{\mathbf{K}}
\usepackage{marginnote}
\newcommand{\mn}[1]{\marginpar{\tiny #1}}

\usepackage{color}

\begin{document}

\title{Peierls Transition in Gross-Neveu Model from Bethe Ansatz}

\author{Valdemar Melin $^a$, Yuta Sekiguchi$^b$, Paul Wiegmann$^c$ and Konstantin Zarembo$^{b,d}$}
\affiliation{$^a$Department of Physics, KTH Royal Institute of Technology,  Stockholm, Sweden}
\affiliation{$^b$Nordita, KTH Royal Institute of Technology and Stockholm University,
Stockholm, Sweden}
\affiliation{$^c$Kadanoff Center for Theoretical Physics, University of Chicago, Chicago, USA}
\affiliation{$^d$Niels Bohr Institute, Copenhagen University, Copenhagen, Denmark}

\begin{abstract}
The two-dimensional Gross-Neveu model is anticipated to undergo a crystalline phase transition at high baryon charge densities. This conclusion is drawn from the mean-field approximation, which closely resembles models of Peierls instability. We demonstrate that this transition indeed occurs when both the rank of the symmetry group and the dimension of the particle representation contributing to the baryon density are large (the
large \(N\)-limit). We derive this result through the exact solution of the model, developing
the large \(N\)-limit of the Bethe Ansatz.  Our analytical construction of the large-$N$ solution of the Bethe Ansatz equations aligns perfectly with the periodic (finite-gap) solution of the Korteweg-de Vries (KdV)  of the mean-field analysis.  
\end{abstract}

\maketitle
The Gross-Neveu (GN) model is a (1+1)-dimensional theory of $N$ interacting Dirac fermions:
\begin{equation}\label{Lagrangian}
 \mathcal{L}=\bar{\psi }_i\left(i\gamma ^\mu \partial _\mu+\mu \gamma ^0 -\sigma \right)\psi _i-\frac{N}{2\lambda }\,\sigma ^2.
\end{equation}
The Hubbard-Stratonovich field $\sigma $ mediates local four-fermion interaction 
$\frac\lambda N(\bar \psi\psi)^2$  (on-shell  $\sigma =-\frac\lambda N \bar{\psi }\psi $).
At zero chemical potential \(\mu\), the  model features asymptotic freedom, spontaneous
breaking of chiral symmetry, and dynamical mass generation: $m= \Lambda \,\lambda ^{1/2(N-1)}{\rm e}\,^{-\pi N/\lambda(N-1) }$  \cite{Gross:1974jv}.
This makes it a compelling case study for non-perturbative effects in quantum field theory. Furthermore, the model is integrable, with a precisely known particle content, spectrum, and scattering S-matrix \cite{Zamolodchikov:1978xm,Karowski:1980kq}.

The model possesses the global $O(2N)\times \mathbbm{Z}_2$ symmetry, where $O(2N)$ acts on a multiplet formed by real and imaginary components of $\psi _i$, and $\mathbbm{Z}_2$ is the chiral symmetry: $\psi _i\rightarrow \gamma ^3\psi _i$, $\sigma \rightarrow -\sigma $. When the fermions are integrated out (at zero $\mu $) the effective potential for $\sigma $ has two minima at $\langle\sigma\rangle =\pm m$ leading to spontaneous breaking of chiral symmetry and fermion mass generation. 

The 
 phase diagram of the model at non-zero chemical potential is quite rich.
Chiral symmetry is restored at high temperatures, whereas at lower temperatures and higher density, the system transitions into a crystalline phase  \cite{Thies:2003kk,Schnetz:2004vr,Thies:2006ti,Basar:2009fg,Ciccone:2022zkg,Ciccone:2023pdk}.
These conclusions are drawn from the large-$N$ approximation, which justifies the mean-field approach. 

The  motivation for our study is  two-fold:
first, to elucidate the crystalline phase through a complete quantum solution using the Bethe-Ansatz technique, and second, to establish a framework for a fully non-perturbative exploration of dense states that is valid a priori for finite $N$ going beyond the mean-field. Furthermore, the large \(N\)-limit of the GN model provides insight into the intricate (or rather singular) relationship between quantum and classical integrable systems.
  
A spontaneous crystalline structure of the  ground state of, otherwise a translational-invariant  fermionic system, is a consequence of the  Peierls instability
\cite{peierls1955quantum,Frhlich1954OnTT}, a celebrated phenomenon extensively studied in the condensed matter literature  (see  \cite{brazovskii1980exact,Horovitz:1981zz,nakahara1981soliton,brazovskii1984electron}
and references therein). In the traditional condensed-matter setting the instability of electronic band structure  is caused by an adibatic  interaction between
lattice phonons and  electrons.    The
 self-interaction in the GN model  \eqref{Lagrangian} mediated by the field
\(\sigma\) yields  the same effect. The large  \(N\)  plays a role of the
adibatic parameter of the Peierles model. Furthermore, the real symmetry group
\(O(2N)\)  gives rise to the {\it commensurate} Peierls transition. When an increase in the chemical potential
pushes the Fermi energy into the conduction zone, the field \(\sigma\) which minimizes  the fermionic energy becomes a spatial-periodic potential with a half-period completely determined by the chemical potential. The energy gain from rearranging the fermion spectrum
 (thankfully to the large \(N\)) exceeds the energy cost of distorting the
environment described by \(\sigma\). In the GN model the instability manifests
itself as a complex pole in the $\sigma $ propagator within a small range
of momenta around  $p\sim 2\mu $ \cite{Koenigstein:2021llr}, pointing towards
transition to a periodically modulated chiral condensate $\left\langle \sigma
\right\rangle$. From a semiclassical perspective the spacially modulated
chiral condensate becomes
 a dominant saddle-point of the large-$N$ path integral \cite{,Basar:2009fg}
above the critical chemical potential   $\mu >2m/\pi$ \cite{Thies:2003kk}.

The mean-field value of \(\sigma\) appears to be  the {\it cnoidal} wave, an exact periodic wave solution of the Korteweg de Vries equation described by Jacobi  elliptic function \(\langle\sigma(x)\rangle=m k^{1/2}{\rm\
sn}(x;k)\).     The cnoidal wave is the simplest
and the best studied instance of a special  class of  potentials known as  finite-gap potential \cite{Novikov:1984id}.   The spectrum of particles  in such potentials possesses only a finite number of gaps.  The   \(E\to -E\) symmetric spectrum of the Dirac equation   in a cnoidal potential possesses  the minimal number, two,    symmetrically situated\mn{}   bandgaps with ends at \((-\varepsilon_+,-\varepsilon_-)\) and \((\varepsilon_-,\varepsilon_+)\). The ends of the spectrum 
are determined solely by the chemical potential in such manner that the Fermi level is located inside one of the gaps.  Furthermore, the density of states in the finite gap potential, referred to as a {\it spectral curve} and the wave functions of the states  are also completely determined by the ends of the spectrum. In this paper we will show how to obtain the spectral curve  as a large \(N\)-limit of the  Bethe Ansatz solution of the full-fledged quantum GN model. 

 The fate of the fermionic 
crystal beyond the semiclassical approximation is an open question. One expects on general grounds that quantum corrections melt the crystal and the long-range order slowly decays at very large distances  \cite{Witten:1978qu}:
\begin{equation}\label{sigma-sigma}
 \left\langle \sigma (x)\sigma (0)\right\rangle
 \sim 
 \frac{\cos \pi \nu x}{|x|^{\frac{\alpha }{N}}}\,,
\end{equation}
due to coupling to the Goldstone mode \cite{berezinskii1971destruction,Kosterlitz:1973xp}. Here $\alpha $ is a parameter of order one. As detailed later, $\nu $ is the density of kinks in the ground state.
This behavior occurs in the  $U(N)$ version of the GN model, solvable exactly by bosonization \cite{Ciccone:2022zkg}. 
The decay of the long-range order is invisible in the strict large-$N$ limit, and the Bethe Ansatz framework developed here may help to clarify the precise nature of the high-density phase. 

The spectrum of the GN model consists of the elementary fermion, its bound states, and solitons.  The solitons transform in the spinor representations of the $D_N$ algebra because kinks of  the $\sigma $-field that interpolate between the two vacua harbour $N$ topologically protected zero modes, one for each fermion flavor, that can be in $2^N$ internal states  \cite{Witten:1977xv}. The  mass spectrum of all particles is known exactly from the factorized scattering matrix  \cite{Zamolodchikov:1978xm}
\begin{equation}\label{masses}
 m_a=m\,\frac{\sin\frac{\pi a}{2N-2}}{\sin\frac{\pi }{2N-2}}\,,\qquad 
 \ms=m_{\bar {\rm s}}=\frac{m}{2\sin\frac{\pi }{2N-2}}\,,
\end{equation}
with $a=1,\ldots,N-2 $ enumerate the fundamental representation of  \(O(2N)\) labelled by nodes of the  \(D_{N}\) Dynkin diagram.   Owing to zero modes, kinks carry \(N/2\) units of the baryon charge \cite{Jackiw:1975fn}. 
%the only meaningful assignment is $-1/2$ for an empty level and $+1/2$ for %an occupied one. 
%The baryon charge of a kink with all levels occupied is thus $N/2$,
The baryon charge of the antisymmetric tensor of rank \(a\) is equal to  \(a\) as it is a  bound state of \(a\) vector particles:
\begin{equation}\label{charges}
 \mathcal{B}_a=a,\qquad \mathcal{B}_{\rm s}=\frac{N}{2}\,.
\end{equation}

Imagine now dispersing some amount $\mathcal{B}=  \sum_a n_a\mathcal{B}_a+\mathcal{B}_{\rm
s} n_{\rm s}$ of baryon charge with occupation numbers \(n_a\) in an otherwise empty system. The ground state energy of such state is just  the  activation energy   \(\sum_a n_a m_a+n_{\rm s} m_{\rm s}\). The  smallest energy could be achieved   by dispersing particles with the  smallest mass-to-charge ratio.  A simple inspection of (\ref{masses}), (\ref{charges}) shows that the optimal choice is kinks!   In spite of their large mass kinks are most energy-efficient   because they can store a large amount of baryon charge in their fermion zero modes.
Kinks start to be created as soon as their chemical potential exceeds the mass: $\mus>\ms$. In terms of the chemical potential of baryons entered in (\ref{Lagrangian}), $\mus=\mu\mathcal{B}_{\rm s}=\mu N/2$. Therefore, the kinks  start to form a   crystalline phase  at $\mu>\mu_c$, where
\begin{equation}\label{mc}
  \mu _c=\frac{\mus}{\mathcal{B}_{\rm s}}=\frac{m}{N\sin\frac{\pi }{2N-2}}\stackrel{N\rightarrow \infty }{\simeq }
  \frac{2m}{\pi }\,.
\end{equation}
These arguments are in 
agreement
with the mean-field analysis  \cite{Thies:2003kk,Schnetz:2004vr,Thies:2006ti,Basar:2009fg}.

The exact structure of the ground state is described by Bethe Ansatz, which we consider here in its thermodynamic form (albeit at zero temperature) focusing on the ground state. As we already explained the ground state is formed by spinors, the kinks. Owing to integrability we are able to characterize these particles by their dispersion \(\varepsilon(p)\) and to uniformize the  dispersion by the   rapidity \((\varepsilon(\theta),\ p(\theta))\), a variable in which the scattering matrix is of  the difference form \(S(\theta-\theta') \).    Furthermore, since the system is  relativistic the dispersion 
is  \((m_{\rm s}\cosh\theta, m_{\rm s}\sinh\theta)\). Then the density of rapidities \(\rho(\theta)d\theta=dp\) of particles  forming the ground states  (kinks, in this case) satisfies a closed equation:
\begin{equation}
 \rho (\theta )-\int\limits_{-B}^{B} \,K(\theta -\eta )\rho (\eta )d\eta=\ms\cosh\theta\,.\label{rho}
\end{equation}
All states within the Fermi interval of rapidity $(-B,B)$ are occupied by the kinks. The value of $B$ is fixed by the condition that the integral of $\rho(\theta)$ over the Fermi interval determines the  total baryon density  $\mathcal{B}/ L=N/2\int^B_{-B}\rho\, d\theta \rho (\theta )$. The value of $\rho(\theta)$ outside the Fermi interval ($|\theta|>B$) defines the momentum of an excited state.
The kernel in the integral equation originates from the scattering phase shift 
$2\pi iK(\theta )=d\ln S(\theta )/d\theta $, where $S(\theta)$ is the kink-to-kink S-matrix restricted to the highest weight state, whose explicit form is presented below.

The energy of the kink, in turn, is given by the equation 
\begin{equation}\label{epsilon-eq}
  \varepsilon  (\theta )-\int\limits_{-B}^{B} \,K(\theta -\eta )\varepsilon  (\eta )d\eta=\ms\cosh\theta-\mus
\end{equation}
conditioned that  the energy $\varepsilon(\theta)$ is negative within the Fermi interval  and positive outside. At the endpoints the energy vanishes:  $\varepsilon (\pm B)=0$, and this condition self-consistently determines the Fermi rapidity. The two equations \eqref{rho} and  \eqref{epsilon-eq} thus form a closed system, refered to as  thermodynamic Bethe Ansatz.   Once the $\varepsilon(\theta)$ and \(B\) are found the energy \( \cal E\) and the  Landau potential \(\Phi=\mathcal{E}-\mu\mathcal{B}\) of the ground state are found by  integration:
\begin{align}\label{free}
&\mathcal{E}=\frac{\ms}{2\pi }\int\limits_{-B}^{B}\rho(\theta
)\cosh\theta d\theta,\
& \Phi=\frac{\ms}{2\pi }\int\limits_{-B}^{B}\varepsilon (\theta )\cosh\theta d\theta \,.
\end{align}

These equations are applicable to any integrable relativistic system. The model-specific data is encoded in the S-matrix, know in the case at hand from \cite{Karowski:1980kq} that gives for the kernel:
\begin{eqnarray}
 K(\theta )&=&\int\limits_{-\infty }^{+\infty }\frac{d\omega }{2\pi }\,\,
 \,{\rm e}\,^{-i\omega \theta }\widetilde{K}(\omega ),
\nonumber \\
\widetilde{K}(\omega )&=& 1-\frac{\,{\rm e}\,^{\frac{\pi |\omega |}{2N-2}}\left(\tanh\frac{\pi \omega }{2}+\tanh\frac{\pi \omega }{2N-2}\right)}{4\sinh\frac{\pi \omega }{2N-2}}\,.
\end{eqnarray}
One can  check some general properties:
the solution exists for any $\mus>\mu_c$, with a positive density $\rho(\theta)$ and the energy $\varepsilon(\theta)$ crossing zero linearly at the endpoints of the Fermi interval. The Landau potential behaves as $\Phi\sim -\delta ^{3/2}$ at the activation threshold: $\delta =\mu /\mu _c-1$. From this we conclude that kinks crystallize at a second-order phase transition.

The BA equations can be easily solved numerically. 
At large \(N\) they can be solved  analytically. 
Similar problems have been studied before, in particular for the GN model with the ground state formed by elementary fermions rather than kinks \cite{Forgacs:1991rs,Chodos:1996pp,Marino:2019eym,DiPietro:2021yxb}.
The interaction kernel is then $\mathcal{O}(1/N)$ and the equations are simply solved by iteration which starts from free particles. A very different situation occurs when the ground state is formed by kinks. In this case,  the interaction is a  dominant effect and the solution is non-trivial already at the leading order in $1/N$. This aspect appears similar to   the
solution of the large-$N$ principal chiral field \cite{Fateev:1994dp,Fateev:1994ai,Kazakov:2023imu},
even if an underlying physics is very different.

The BA equations for GN kinks at a small $N$ were studied using perturbative techniques: for  $N=2$  \cite{DiPietro:2021yxb} and for $N=4$. The latter
case, the $O(8)$ model, is special due to the external  automorphisms, a `triality', making the vector and the spinor particles equivalent. The  analysis of the  BA equations of Ref.  \cite{Marino:2019eym,Marino:2023epd} then also applies to kinks upon substitution $\mu \rightarrow \mus=2\mu $.

We will be able to solve the BA equations because at  large-$N$ the scattering phase  becomes singular and is of the order of $\mathcal{O}(N)$:
\begin{equation}\label{K-tilde-large-N}
 \widetilde{K}(\omega )\stackrel{N\rightarrow \infty }{\simeq } -N\, \frac{\tanh\frac{\pi \omega }{2}}{2\pi \omega }\,.
\end{equation}
The first term   in the LHS of (\ref{epsilon-eq}),    representing the
energy of spinors outside of the Fermi interval $(-B,B)$ is then negligible, being much smaller than the scattering phase.  
Converting (\ref{K-tilde-large-N}) to the $\theta $-representation we obtain a singular integral equation
\begin{equation}\label{BAE-epsx}
\frac{1}{4\pi ^2}\int\limits_{-B}^{B}d\eta \,\varepsilon (\eta )\ln\coth^2\frac{\theta-\eta  }{2}=\frac{m}{\pi }\cosh\theta -\frac{\mu }{2}\,.
\end{equation}
Upon differentiation in $\theta $ it takes the standard form with the $1/\sinh$ kernel and can be solved by standard techniques \cite{Gakhov}.
The solution that goes to zero at the endpoints is 
\begin{equation}\label{eps-solxx}
  \varepsilon (\theta )=-2m\sqrt{\sinh^2B-\sinh^2\theta }\,.
\end{equation}

This cannot be the end of the story because the BA equation and the condition $\varepsilon (B)=0$ should also fix $B$. This extra condition was lost upon differentiation. Going back to the original equation with the logarithmic kernel gives one extra constraint:
\begin{equation}\label{mu->B}
 \frac{\pi \mu }{2m}=\frac{\E}{k}\,,
 \qquad
 k=\frac{1}{\cosh B}\,,
\end{equation}
where $\E\equiv \E(k) $ is the complete elliptic integral of the second kind (in the notations of \cite{gradshteyn2014table}). This condition determines the Fermi rapidity $B$ as a function of the chemical potential. Integrating (\ref{free}) we get the free energy:
\begin{equation}\label{free-lunch}
 \Phi =-\frac{Nm^2\sinh^2B}{2\pi }\,,
\end{equation}
or, equivalently,
\begin{equation}
    \mathcal{E}-\mu_{\rm s} \nu=-\frac {N m^2 k'{}^2}{\pi k^2}\,,
\end{equation}
where $\nu$ is the density of kinks and $k'=\sqrt{1-k^2}=\tanh B$ is the complementary modulus of the elliptic integral.

At large densities, $\mu \gg m$, the Fermi rapidity grows logarithmically:
$ B\simeq \ln(2\mu /{m})$, and $\Phi \simeq -N\mu ^2/2\pi $ as expected for a gas of $N$ species free fermions.  This can be made more precise. Following \cite{Fateev:1994ai} we identify $B$ with the inverse of the running coupling:  
$\lambda (\mu )\equiv \pi /{B}$. Indeed, the $\beta $-function obtained by differentiating (\ref{mu->B}) in $\ln \mu $ then coincides with the one-loop exact large-$N$ expression  up to (scheme-dependent) non-perturbative terms: $\beta =-\frac{\lambda ^2}{\pi }(1+{\rm series~in~}\,{\rm e}\,^{-4\pi /\lambda })$.  This makes $B$ a useful measure of the interaction strength: the system is weakly coupled at asymptotically large densities and becomes strongly when the Fermi interval shrinks to zero size: $B\sim \sqrt{4\delta /|\ln \delta |}$, implying $\Phi \sim -\delta /|\ln\delta |$. This seems to contradict our earlier conclusion that $\Phi \sim -\delta ^{3/2}$. But the large-$N$ approximation breaks down  at $\delta \sim 1/N$ (all three terms in the BA equations are then of the same order). There are thus two regimes at large-$N$: the logarithmic scaling $\delta /|\ln \delta |$  further away from the critical point gives way to a milder $\delta ^{3/2}$ behavior parametrically close to it.

The solution (\ref{eps-solxx}) defines the energy of holes.
To find their dispersion we need to compute the density and then the momentum. In the large-$N$ approximation,
\begin{equation}
\frac{1}{4\pi ^2}\int\limits_{-B}^{B}d\eta \,\rho (\eta )\ln\coth^2\frac{\theta-\eta  }{2}=\frac{m}{\pi }\cosh\theta,
\end{equation}
which again is solved by differentiating in $\theta $.
To maintain positivity of the density we need to allow for an inverse square root, a zero mode of the $1/\sinh$ kernel:
\begin{align}\label{density-sol}
  \rho (\theta) =\frac{m\sinh 2B}{c_s\sqrt{\sinh^2 B - \sinh^2 \theta}}
    -2m\sqrt{\sinh^2 B - \sinh^2 \theta}.
\end{align}
At finite $N$ the density of particles and the density of exited states are smoothly glued at the edge of the interval $(-B,B)$.  The singular behavior we found here is an artifact of the large $N$ limit.

 The coefficient $c_s$ is fixed by substituting the solution back into the  equation with the log-kernel:
\begin{equation}\label{sound}
c_s=\frac{k'\K}{\E}
\end{equation}
where $\K$ is the complete elliptic integral of the first kind with the same $1/\cosh B$ modulus.

For the density of kinks we thus get:
\begin{equation}\label{density}
 \nu \equiv \int\limits_{-B}^{B}\frac{d\theta}{2\pi } \,\rho (\theta )=\frac{m }{k\K}=\frac{2\varepsilon _+\varepsilon _-}{\pi c_s\mu },
\end{equation}
where we used (\ref{mu->B}), (\ref{sound}) to get rid of ellipticae and introduced the notations:
\begin{equation}
 \varepsilon _+=m\cosh B,\qquad \varepsilon _-=m\sinh B.
\end{equation}
As we shall see these parameters characterize the gap in the fermion spectrum.

The parameter $\nu $ plays a dual role. On the one hand, it defines the number density of kinks, how many of them fit in a unit of length, on the other hand, it also defines the largest momentum of a hole: because $dp/d\theta =\rho $, the latter varies between $-\pi \nu$ and $\pi \nu $.

\begin{figure}
    \centering
    \begin{subfigure}[b]{0.55\textwidth}
   \includegraphics[scale=0.9]{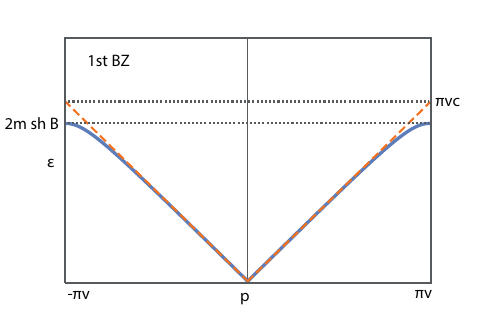}
   \caption{}
      \label{fig:disp-holes} 
\end{subfigure}
 \begin{subfigure}[b]{0.55\textwidth}
    \includegraphics[scale=0.3]{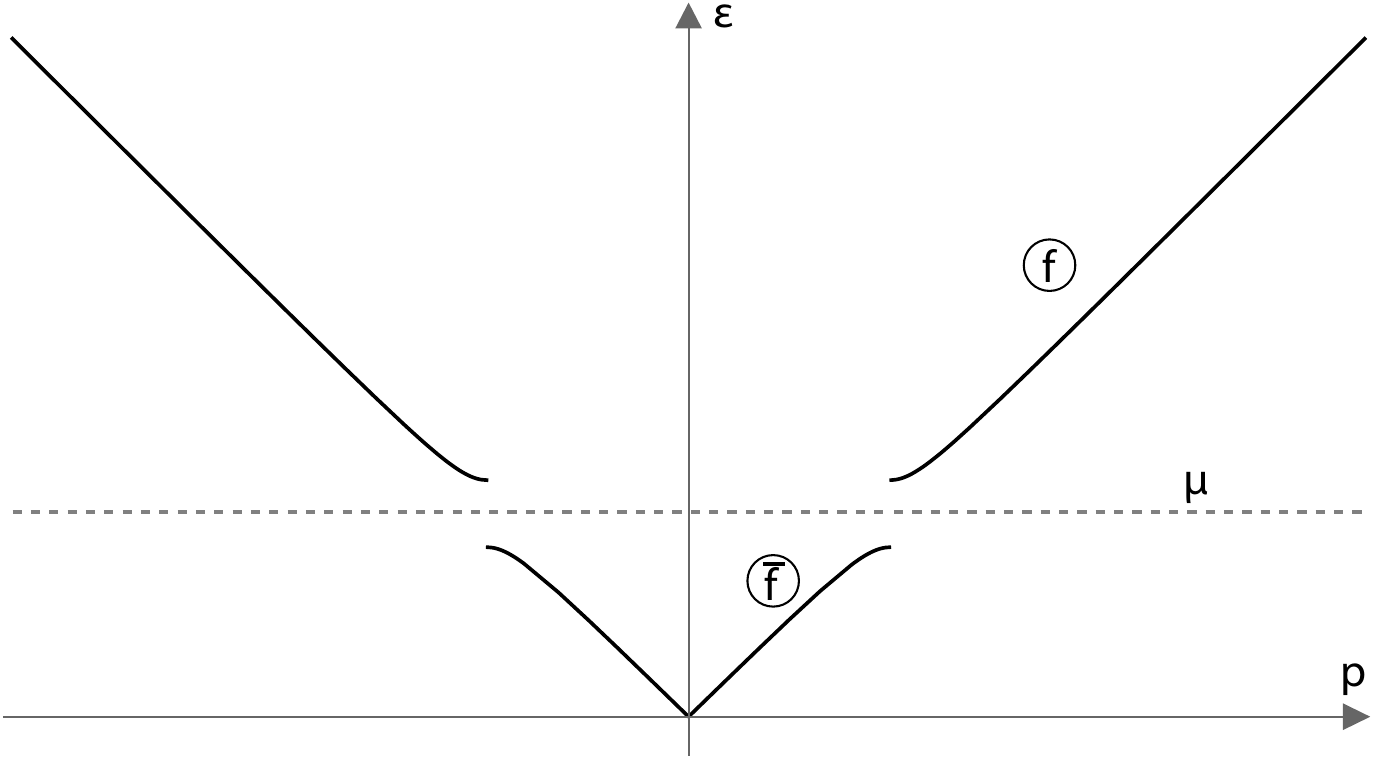}
   \caption{}
      \label{fig:disp-ferms} 
\end{subfigure}
    \label{fig:disp} 
    \caption{The dispersion curves: (a) for the phonon; (b) the two branches of the spectral curve: the upper branch is the spectrum of vector particles (elementary fermions), the lower branch is the spectrum of holes described by kinks of opposite chirality. }
\end{figure}

\begin{figure}
\centering
   \includegraphics[scale=0.6]{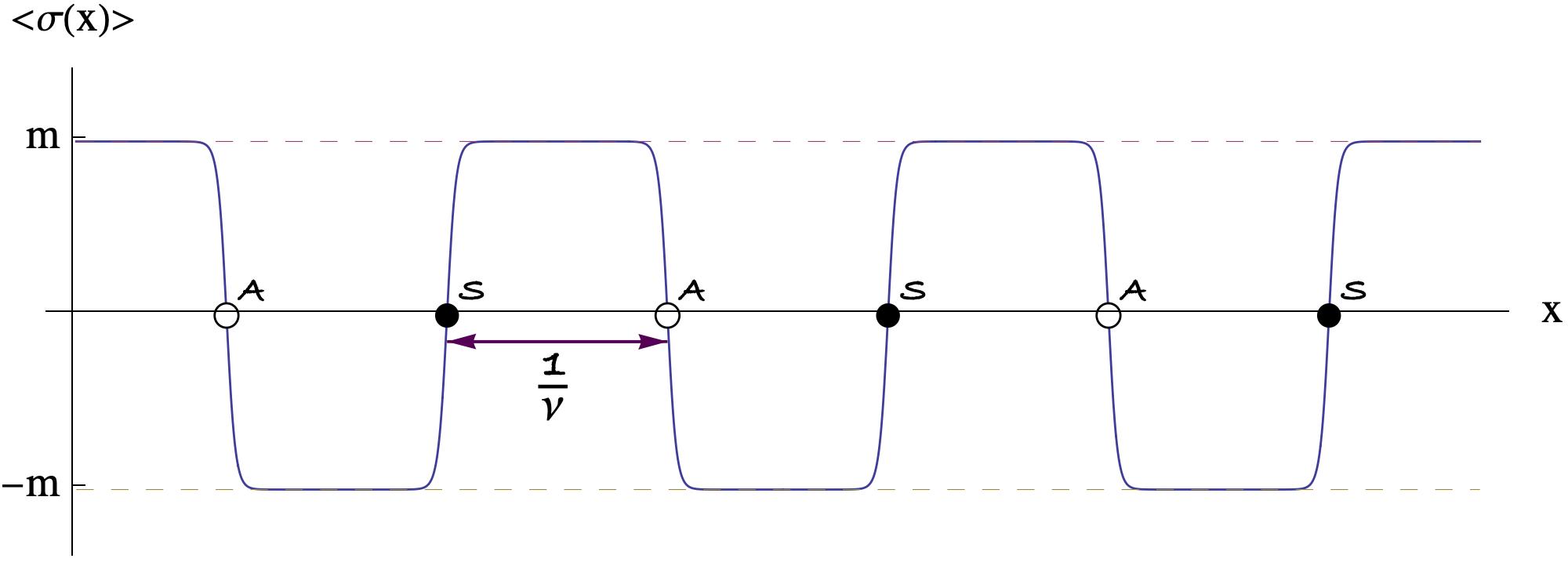}
\caption{The ground state as a periodic soliton.}
  \label{fig:Ng2}
\end{figure}

We can find the dispersion relation of a hole by combining (\ref{eps-solxx}) with (\ref{density-sol}):
\begin{equation}\label{diff-disp}
 c_sdp=\frac{4\varepsilon _+\varepsilon _--c_s\varepsilon ^2}{\sqrt{\left(4\varepsilon _+^2-\varepsilon ^2\right)
 \left(4\varepsilon _-^2-\varepsilon ^2\right)}}\,d\varepsilon .
\end{equation}
The curve $\varepsilon (p)$ is depicted in fig.~\ref{fig:disp-holes}. The dispersion is approximately linear at small momenta:
\begin{equation}
 \varepsilon \stackrel{p\rightarrow 0}{\simeq }c_s|p|,
\end{equation}
as expected of the sound mode. All this suggests to identify a hole in the distribution of kinks with a phonon, the vibrational mode of the chiral crystal. The ground state can be pictured as a collection of kinks placed equidistantly as illustrated in fig.~\ref{fig:Ng2}. Removing one kink is equivalent to sending an acoustic wave across the lattice. Since the soliton centres are separated by $1/\nu$, the lattice vibrations naturally fit into a Brillouin zone $-\pi \nu<p<\pi \nu  $, and so do the holes in the BA. 

The sonic nature of the holes becomes particularly lucid at small $B$ when $\varepsilon _+\gg\varepsilon _-, \varepsilon $, and the dispersion curve simplifies:
\begin{equation}
 c_sdp\stackrel{\mu \rightarrow \mu _c}{\simeq }\frac{2\varepsilon _-d\varepsilon }{\sqrt{4\varepsilon _-^2-\varepsilon ^2}}\,.
\end{equation}
Taking into account that $\varepsilon _+\simeq m\simeq \pi \mu /2$ and expressing $\varepsilon _-$ from (\ref{density}), we find:
\begin{equation}
 \varepsilon \stackrel{\mu \rightarrow \mu _c}{\simeq } 2c_s\nu \left|\sin\frac{p}{2\nu }\right|,
\end{equation}
the dispersion of a phonon in a harmonic lattice with spacing $1/\nu $. The energy is a periodic function of momentum with a period $2\pi \nu $ not only in this approximation but actually for any $B$.  
The speed of sound experiences a critical slowdown at small $B$:  $c_s\simeq B\ln(4/B)\sim \sqrt{\delta |\ln \delta |}$,  grows with $\mu $, and approaches the speed of light in the perturbative regime. 

Since
the soliton lattice contains two particles per unit cell (fig.~\ref{fig:Ng2}), the fermion Brillouin zone is half that of phonons: 
$-\pi \nu/2<p< \pi \nu /2$. This fits well with the mechanics of Peierls instability  \cite{peierls1955quantum}, indeed at weak coupling (large $\mu $) $\pi \nu \simeq 2\mu $ and the band gap opens exactly at the Fermi level. Once interactions become stronger the density of kinks $\nu $ and with it the size of the Brillouin zone diminishes faster than $\mu $. At the critical point ($\mu\rightarrow \mu _c$) the density vanishes as $\nu \sim 1/|\ln\delta |$. 

Now we turn to the part of the spectrum which corresponds to vector particles, the quanta of the fermion field in the Lagrangian \eqref{Lagrangian}. They do form a Fermi interval of their own, their exact energy is overall positive and is determined by the following BA equation:
\begin{equation}\label{fermion-BAE}
  \varepsilon_{{\rm f}}(\theta )-\int\limits_{-B}^{B}d\eta \,K_{\rm f}(\theta -\eta )\varepsilon  (\eta )=m\cosh\theta-\mu .
\end{equation}
There is a similar equation for $dp_{\rm f}/d\theta $. In both of these equations
the kernel is the derivative of the fermion-kink scattering phase \cite{Karowski:1980kq}:
\begin{equation}
 \widetilde{K}_{\rm f}(\omega )=-\frac{\,{\rm e}\,^{\frac{\pi |\omega |}{2N-2}}}{2\cosh\frac{\pi \omega }{2}}\,
 \qquad 
 K_{\rm f}(\theta )\stackrel{N\rightarrow \infty }{\simeq }-\frac{1}{2\pi \cosh\theta }\,.
\end{equation}
Solving the equation at large-$N$ we find the spectral curve of an elementary fermion:
\begin{eqnarray}
&& \varepsilon _{\rm f}=m\sqrt{\sinh^2 B + \cosh^2\theta} - \mu
\vphantom{\frac{dp_{\rm f}}{d\theta }=  m\sqrt{\sinh^2 B + \cosh^2 \theta}-\frac{m\sinh 2B}{2c_s\sqrt{\sinh^2 B + \cosh^2 \theta}}}
 \\
 \nonumber
 && \frac{dp_{\rm f}}{d\theta }=  m\sqrt{\sinh^2 B + \cosh^2 \theta}-\frac{m\sinh 2B}{2c_s\sqrt{\sinh^2 B + \cosh^2 \theta}}\,.
\end{eqnarray}
Excluding $\theta $ gives a differential equation:
\begin{equation}\label{alg-curve}
 c_sdp=\frac{c_s\epsilon ^2-\varepsilon _+\varepsilon _-}{
 \sqrt{\left(\epsilon ^2-\varepsilon _+^2\right)\left(\epsilon ^2-\varepsilon _-^2\right)}}d\epsilon,\qquad \epsilon =\varepsilon _{\rm f}+\mu . 
\end{equation}

Apart from fermions we also expect to find hole-type excitations obtained by removing one fermion from the ground state. This leaves one kink with $N-1$ out of $N$ levels filled, changing the chirality of its $O(2N)$ spinor representation. By some abuse of terminology we call kinks with one empty level anti-kinks. The kernel of their BA equation is determined by the scattering phase for $O(2N)$ spinor representations of opposite chirality \cite{Karowski:1980kq}:
\begin{equation}
 K_{\bar{{\rm f}}}(\omega )=K(\omega )-\frac{1}{\,{\rm e}\,^{\frac{\pi |\omega |}{N-1}}+1},
~
 K_{\bar{{\rm f}}}(\theta )\stackrel{N\rightarrow \infty }{\simeq }K(\theta )-\frac{\delta (\theta )}{2}.
\end{equation}
Using the BA equation for the background (\ref{epsilon-eq}) and taking into account that $\mu_{\bar{{\rm f}}}=\mu _{\rm s}-\mu $ we find:
\begin{equation}
 \varepsilon_{\bar{{\rm f}}}(\theta )=\mu +\frac{1}{2}\,\varepsilon (\theta ),
 \qquad 
 p_{\bar{{\rm f}}}(\theta )=\frac{1}{2}\,p(\theta ),
\end{equation}
for $|\theta |<B$. The energy of kinks (\ref{eps-solxx}) is negative inside this rapidity interval but the upward shift by $\mu $ makes the energy of anti-kinks overall positive.

Their spectral curve  follows from (\ref{diff-disp}) which, as can be easily seen, implies that the energy of anti-kinks satisfies the same equation (\ref{alg-curve}) as the energy of fermions  if we now set $\epsilon =\mu -\varepsilon_{\bar{{\rm f}}}$ and choose the different branch of the square root. The two branches are shown in fig.~\ref{fig:disp-ferms} and together form the expected spectral curve of a Dirac fermion on the background of the periodic soliton.

It is quite remarkable that in the large $N$ limit the fermion and anti-kink energies are different branches of the same elliptic spectral curve. This fact is the main aspect of the Peierls phenomenon and solely relies on the theory of periodic solutions of classical soliton equations \cite{Novikov:1984id}. Moreover it can be identified with the spectral curve of the mKdV equation \cite{brazovskii1980exact} whose periodic soliton solution describes the background chiral crystal $\left\langle \sigma (x)\right\rangle$  \cite{brazovskii1980exact,brazovskii1984electron}, thus establishing direct contact between the BA and the equations of the mean-field theory. The elliptic integrals we encountered before are defined on the same elliptic curve, and one can check that the free energy (\ref{free-lunch}) and the chemical potential (\ref{mu->B}) agree precisely with the predictions of the mean-field theory \cite{brazovskii1984electron}.

The common dispersion relation for fermions and holes starts linearly at small $p$ and has a bandgap at the boundary of the Brillouin zone $p=\pi \nu /2$. Indeed, the smallest fermion energy is $\varepsilon _{\rm f}(\theta =0)=\varepsilon _+-\mu $, while the smallest hole energy is $\varepsilon _{\bar{{\rm f}}}(\theta =0)=\mu -\varepsilon _-$. The gap equals their sum:
\begin{align}
 \Delta = \varepsilon _+-\varepsilon _-=m\,{\rm e}\,^{-B}.
\end{align}
The gap is finite at the critical point: $\Delta(\mu _c) =m$, diminishes with $\mu $, and becomes very small at weak coupling: $\Delta \simeq m^2/(2\mu )\sim \mu \,{\rm e}\,^{-2\pi /\lambda }$. This is fully non-perturbative and cannot be detected at any order of the weak-coupling expansion.

We have also solved the BA equations numerically at finite $N$. One conclusion we can draw from these preliminary studies is that the mean-field predictions are strikingly robust. We can confirm that (i) the GN model undergoes a second-order Peierls transition at any $N\geq 2$. Once the transition is identified with the creation threshold of kinks, the critical chemical potential (\ref{mc}) follows from the exact mass formula. This is an exact result that does not rely on the large-$N$ approximation. (ii) The high-density phase is characterized by a quasi-long-range crystalline order (\ref{sigma-sigma}). Since
the Bethe wavefunction of the ground state is translationally-invariant by construction, translational symmetry is not broken. (iii) The spectrum contains a gapless phonon (also at finite $N$) described non-perturbatively as a hole in the Fermi sea of kinks; (iv) the fermion spectrum remains gapped at any $N$ and any coupling strength. Our preliminary numerical studies show that $1/N$ corrections are significant and reach up to $10\%$ even at $N=\mathcal{O}(100)$. More details on the full quantum solution of the model will be given in a separate publication. 

\paragraph{Acknowledgements.} We would like to thank Z. Bajnok, A. Balatsky, J. Balog, L. Di Pietro, V. Kazakov, D. Kuzmanovski, E. Langmann, O. Ohlsson Sax, M. Serone, F. Smirnov, E. Sobko and D. Volin for illuminating discussions. P.W. acknowledges helpful discussions with N. Kirova and S. Brazovski. The work of P.W. was supported by the NSF under Grant NSF DMR-1949963. The work of K.Z. was supported by VR grant 2021-04578. Nordita was partially supported by NordForsk.


\begin{thebibliography}{32}%
\makeatletter
\providecommand \@ifxundefined [1]{%
 \@ifx{#1\undefined}
}%
\providecommand \@ifnum [1]{%
 \ifnum #1\expandafter \@firstoftwo
 \else \expandafter \@secondoftwo
 \fi
}%
\providecommand \@ifx [1]{%
 \ifx #1\expandafter \@firstoftwo
 \else \expandafter \@secondoftwo
 \fi
}%
\providecommand \natexlab [1]{#1}%
\providecommand \enquote  [1]{``#1''}%
\providecommand \bibnamefont  [1]{#1}%
\providecommand \bibfnamefont [1]{#1}%
\providecommand \citenamefont [1]{#1}%
\providecommand \href@noop [0]{\@secondoftwo}%
\providecommand \href [0]{\begingroup \@sanitize@url \@href}%
\providecommand \@href[1]{\@@startlink{#1}\@@href}%
\providecommand \@@href[1]{\endgroup#1\@@endlink}%
\providecommand \@sanitize@url [0]{\catcode `\\12\catcode `\$12\catcode
  `\&12\catcode `\#12\catcode `\^12\catcode `\_12\catcode `\%12\relax}%
\providecommand \@@startlink[1]{}%
\providecommand \@@endlink[0]{}%
\providecommand \url  [0]{\begingroup\@sanitize@url \@url }%
\providecommand \@url [1]{\endgroup\@href {#1}{\urlprefix }}%
\providecommand \urlprefix  [0]{URL }%
\providecommand \Eprint [0]{\href }%
\providecommand \doibase [0]{http://dx.doi.org/}%
\providecommand \selectlanguage [0]{\@gobble}%
\providecommand \bibinfo  [0]{\@secondoftwo}%
\providecommand \bibfield  [0]{\@secondoftwo}%
\providecommand \translation [1]{[#1]}%
\providecommand \BibitemOpen [0]{}%
\providecommand \bibitemStop [0]{}%
\providecommand \bibitemNoStop [0]{.\EOS\space}%
\providecommand \EOS [0]{\spacefactor3000\relax}%
\providecommand \BibitemShut  [1]{\csname bibitem#1\endcsname}%
\let\auto@bib@innerbib\@empty
%</preamble>
\bibitem [{\citenamefont {Gross}\ and\ \citenamefont
  {Neveu}(1974)}]{Gross:1974jv}%
  \BibitemOpen
  \bibfield  {author} {\bibinfo {author} {\bibfnamefont {D.~J.}\ \bibnamefont
  {Gross}}\ and\ \bibinfo {author} {\bibfnamefont {A.}~\bibnamefont {Neveu}},\
  }\href {\doibase 10.1103/PhysRevD.10.3235} {\bibfield  {journal} {\bibinfo
  {journal} {Phys. Rev. D}\ }\textbf {\bibinfo {volume} {10}},\ \bibinfo
  {pages} {3235} (\bibinfo {year} {1974})}\BibitemShut {NoStop}%
\bibitem [{\citenamefont {Zamolodchikov}\ and\ \citenamefont
  {Zamolodchikov}(1979)}]{Zamolodchikov:1978xm}%
  \BibitemOpen
  \bibfield  {author} {\bibinfo {author} {\bibfnamefont {A.~B.}\ \bibnamefont
  {Zamolodchikov}}\ and\ \bibinfo {author} {\bibfnamefont {A.~B.}\ \bibnamefont
  {Zamolodchikov}},\ }\href {\doibase 10.1016/0003-4916(79)90391-9} {\bibfield
  {journal} {\bibinfo  {journal} {Annals Phys.}\ }\textbf {\bibinfo {volume}
  {120}},\ \bibinfo {pages} {253} (\bibinfo {year} {1979})}\BibitemShut
  {NoStop}%
%%CITATION = APNYA,120,253;%%
\bibitem [{\citenamefont {Karowski}\ and\ \citenamefont
  {Thun}(1981)}]{Karowski:1980kq}%
  \BibitemOpen
  \bibfield  {author} {\bibinfo {author} {\bibfnamefont {M.}~\bibnamefont
  {Karowski}}\ and\ \bibinfo {author} {\bibfnamefont {H.~J.}\ \bibnamefont
  {Thun}},\ }\href {\doibase 10.1016/0550-3213(81)90484-3} {\bibfield
  {journal} {\bibinfo  {journal} {Nucl. Phys. B}\ }\textbf {\bibinfo {volume}
  {190}},\ \bibinfo {pages} {61} (\bibinfo {year} {1981})}\BibitemShut
  {NoStop}%
\bibitem [{\citenamefont {Thies}\ and\ \citenamefont
  {Urlichs}(2003)}]{Thies:2003kk}%
  \BibitemOpen
  \bibfield  {author} {\bibinfo {author} {\bibfnamefont {M.}~\bibnamefont
  {Thies}}\ and\ \bibinfo {author} {\bibfnamefont {K.}~\bibnamefont
  {Urlichs}},\ }\href {\doibase 10.1103/PhysRevD.67.125015} {\bibfield
  {journal} {\bibinfo  {journal} {Phys. Rev. D}\ }\textbf {\bibinfo {volume}
  {67}},\ \bibinfo {pages} {125015} (\bibinfo {year} {2003})},\ \Eprint
  {http://arxiv.org/abs/hep-th/0302092} {arXiv:hep-th/0302092} \BibitemShut
  {NoStop}%
\bibitem [{\citenamefont {Schnetz}\ \emph {et~al.}(2004)\citenamefont
  {Schnetz}, \citenamefont {Thies},\ and\ \citenamefont
  {Urlichs}}]{Schnetz:2004vr}%
  \BibitemOpen
  \bibfield  {author} {\bibinfo {author} {\bibfnamefont {O.}~\bibnamefont
  {Schnetz}}, \bibinfo {author} {\bibfnamefont {M.}~\bibnamefont {Thies}}, \
  and\ \bibinfo {author} {\bibfnamefont {K.}~\bibnamefont {Urlichs}},\ }\href
  {\doibase 10.1016/j.aop.2004.06.009} {\bibfield  {journal} {\bibinfo
  {journal} {Annals Phys.}\ }\textbf {\bibinfo {volume} {314}},\ \bibinfo
  {pages} {425} (\bibinfo {year} {2004})},\ \Eprint
  {http://arxiv.org/abs/hep-th/0402014} {arXiv:hep-th/0402014} \BibitemShut
  {NoStop}%
\bibitem [{\citenamefont {Thies}(2006)}]{Thies:2006ti}%
  \BibitemOpen
  \bibfield  {author} {\bibinfo {author} {\bibfnamefont {M.}~\bibnamefont
  {Thies}},\ }\href {\doibase 10.1088/0305-4470/39/41/S04} {\bibfield
  {journal} {\bibinfo  {journal} {J. Phys. A}\ }\textbf {\bibinfo {volume}
  {39}},\ \bibinfo {pages} {12707} (\bibinfo {year} {2006})},\ \Eprint
  {http://arxiv.org/abs/hep-th/0601049} {arXiv:hep-th/0601049} \BibitemShut
  {NoStop}%
\bibitem [{\citenamefont {Basar}\ \emph {et~al.}(2009)\citenamefont {Basar},
  \citenamefont {Dunne},\ and\ \citenamefont {Thies}}]{Basar:2009fg}%
  \BibitemOpen
  \bibfield  {author} {\bibinfo {author} {\bibfnamefont {G.}~\bibnamefont
  {Basar}}, \bibinfo {author} {\bibfnamefont {G.~V.}\ \bibnamefont {Dunne}}, \
  and\ \bibinfo {author} {\bibfnamefont {M.}~\bibnamefont {Thies}},\ }\href
  {\doibase 10.1103/PhysRevD.79.105012} {\bibfield  {journal} {\bibinfo
  {journal} {Phys. Rev. D}\ }\textbf {\bibinfo {volume} {79}},\ \bibinfo
  {pages} {105012} (\bibinfo {year} {2009})},\ \Eprint
  {http://arxiv.org/abs/0903.1868} {arXiv:0903.1868 [hep-th]} \BibitemShut
  {NoStop}%
\bibitem [{\citenamefont {Ciccone}\ \emph {et~al.}(2022)\citenamefont
  {Ciccone}, \citenamefont {Di~Pietro},\ and\ \citenamefont
  {Serone}}]{Ciccone:2022zkg}%
  \BibitemOpen
  \bibfield  {author} {\bibinfo {author} {\bibfnamefont {R.}~\bibnamefont
  {Ciccone}}, \bibinfo {author} {\bibfnamefont {L.}~\bibnamefont {Di~Pietro}},
  \ and\ \bibinfo {author} {\bibfnamefont {M.}~\bibnamefont {Serone}},\ }\href
  {\doibase 10.1103/PhysRevLett.129.071603} {\bibfield  {journal} {\bibinfo
  {journal} {Phys. Rev. Lett.}\ }\textbf {\bibinfo {volume} {129}},\ \bibinfo
  {pages} {071603} (\bibinfo {year} {2022})},\ \Eprint
  {http://arxiv.org/abs/2203.07451} {arXiv:2203.07451 [hep-th]} \BibitemShut
  {NoStop}%
\bibitem [{\citenamefont {Ciccone}\ \emph {et~al.}(2024)\citenamefont
  {Ciccone}, \citenamefont {Di~Pietro},\ and\ \citenamefont
  {Serone}}]{Ciccone:2023pdk}%
  \BibitemOpen
  \bibfield  {author} {\bibinfo {author} {\bibfnamefont {R.}~\bibnamefont
  {Ciccone}}, \bibinfo {author} {\bibfnamefont {L.}~\bibnamefont {Di~Pietro}},
  \ and\ \bibinfo {author} {\bibfnamefont {M.}~\bibnamefont {Serone}},\ }\href
  {\doibase 10.1007/JHEP02(2024)211} {\bibfield  {journal} {\bibinfo  {journal}
  {JHEP}\ }\textbf {\bibinfo {volume} {02}},\ \bibinfo {pages} {211} (\bibinfo
  {year} {2024})},\ \Eprint {http://arxiv.org/abs/2312.13756} {arXiv:2312.13756
  [hep-th]} \BibitemShut {NoStop}%
\bibitem [{\citenamefont {Peierls}(1955)}]{peierls1955quantum}%
  \BibitemOpen
  \bibfield  {author} {\bibinfo {author} {\bibfnamefont {R.~E.}\ \bibnamefont
  {Peierls}},\ }\href@noop {} {\emph {\bibinfo {title} {Quantum theory of
  solids}}}\ (\bibinfo  {publisher} {Oxford University Press},\ \bibinfo {year}
  {1955})\BibitemShut {NoStop}%
\bibitem [{\citenamefont {Fr{\"o}hlich}(1954)}]{Frhlich1954OnTT}%
  \BibitemOpen
  \bibfield  {author} {\bibinfo {author} {\bibfnamefont {H.}~\bibnamefont
  {Fr{\"o}hlich}},\ }\href {https://api.semanticscholar.org/CorpusID:122157741}
  {\bibfield  {journal} {\bibinfo  {journal} {Proc. R. Soc.}\ }\textbf
  {\bibinfo {volume} {A223}},\ \bibinfo {pages} {296 } (\bibinfo {year}
  {1954})}\BibitemShut {NoStop}%
\bibitem [{\citenamefont {Brazovskii}\ \emph {et~al.}(1980)\citenamefont
  {Brazovskii}, \citenamefont {Gordyunin},\ and\ \citenamefont
  {Kirova}}]{brazovskii1980exact}%
  \BibitemOpen
  \bibfield  {author} {\bibinfo {author} {\bibfnamefont {S.}~\bibnamefont
  {Brazovskii}}, \bibinfo {author} {\bibfnamefont {S.}~\bibnamefont
  {Gordyunin}}, \ and\ \bibinfo {author} {\bibfnamefont {N.}~\bibnamefont
  {Kirova}},\ }\href@noop {} {\bibfield  {journal} {\bibinfo  {journal} {JETP
  Lett.}\ }\textbf {\bibinfo {volume} {31}},\ \bibinfo {pages} {456} (\bibinfo
  {year} {1980})}\BibitemShut {NoStop}%
\bibitem [{\citenamefont {Horovitz}(1981)}]{Horovitz:1981zz}%
  \BibitemOpen
  \bibfield  {author} {\bibinfo {author} {\bibfnamefont {B.}~\bibnamefont
  {Horovitz}},\ }\href {\doibase 10.1103/PhysRevLett.46.742} {\bibfield
  {journal} {\bibinfo  {journal} {Phys. Rev. Lett.}\ }\textbf {\bibinfo
  {volume} {46}},\ \bibinfo {pages} {742} (\bibinfo {year} {1981})}\BibitemShut
  {NoStop}%
\bibitem [{\citenamefont {Nakahara}\ and\ \citenamefont
  {Maki}(1981)}]{nakahara1981soliton}%
  \BibitemOpen
  \bibfield  {author} {\bibinfo {author} {\bibfnamefont {M.}~\bibnamefont
  {Nakahara}}\ and\ \bibinfo {author} {\bibfnamefont {K.}~\bibnamefont
  {Maki}},\ }\href@noop {} {\bibfield  {journal} {\bibinfo  {journal} {Phys.
  Rev. B}\ }\textbf {\bibinfo {volume} {24}},\ \bibinfo {pages} {1045}
  (\bibinfo {year} {1981})}\BibitemShut {NoStop}%
\bibitem [{\citenamefont {Brazovskii}\ and\ \citenamefont
  {Kirova}(1984)}]{brazovskii1984electron}%
  \BibitemOpen
  \bibfield  {author} {\bibinfo {author} {\bibfnamefont {S.}~\bibnamefont
  {Brazovskii}}\ and\ \bibinfo {author} {\bibfnamefont {N.}~\bibnamefont
  {Kirova}},\ }\href@noop {} {\bibfield  {journal} {\bibinfo  {journal} {Sov.
  Sci. Rev. A}\ }\textbf {\bibinfo {volume} {5}},\ \bibinfo {pages} {99}
  (\bibinfo {year} {1984})}\BibitemShut {NoStop}%
\bibitem [{\citenamefont {Koenigstein}\ \emph {et~al.}(2022)\citenamefont
  {Koenigstein}, \citenamefont {Pannullo}, \citenamefont {Rechenberger},
  \citenamefont {Steil},\ and\ \citenamefont {Winstel}}]{Koenigstein:2021llr}%
  \BibitemOpen
  \bibfield  {author} {\bibinfo {author} {\bibfnamefont {A.}~\bibnamefont
  {Koenigstein}}, \bibinfo {author} {\bibfnamefont {L.}~\bibnamefont
  {Pannullo}}, \bibinfo {author} {\bibfnamefont {S.}~\bibnamefont
  {Rechenberger}}, \bibinfo {author} {\bibfnamefont {M.~J.}\ \bibnamefont
  {Steil}}, \ and\ \bibinfo {author} {\bibfnamefont {M.}~\bibnamefont
  {Winstel}},\ }\href {\doibase 10.1088/1751-8121/ac820a} {\bibfield  {journal}
  {\bibinfo  {journal} {J. Phys. A}\ }\textbf {\bibinfo {volume} {55}},\
  \bibinfo {pages} {375402} (\bibinfo {year} {2022})},\ \Eprint
  {http://arxiv.org/abs/2112.07024} {arXiv:2112.07024 [hep-ph]} \BibitemShut
  {NoStop}%
\bibitem [{\citenamefont {Novikov}\ \emph {et~al.}(1984)\citenamefont
  {Novikov}, \citenamefont {Manakov}, \citenamefont {Pitaevsky},\ and\
  \citenamefont {Zakharov}}]{Novikov:1984id}%
  \BibitemOpen
  \bibfield  {author} {\bibinfo {author} {\bibfnamefont {S.}~\bibnamefont
  {Novikov}}, \bibinfo {author} {\bibfnamefont {S.~V.}\ \bibnamefont
  {Manakov}}, \bibinfo {author} {\bibfnamefont {L.~P.}\ \bibnamefont
  {Pitaevsky}}, \ and\ \bibinfo {author} {\bibfnamefont {V.~E.}\ \bibnamefont
  {Zakharov}},\ }\href@noop {} {\emph {\bibinfo {title} {Theory of Solitons.
  The Inverse Scattering Method}}}\ (\bibinfo  {publisher} {Consultants
  Bureau},\ \bibinfo {address} {New York, USA},\ \bibinfo {year}
  {1984})\BibitemShut {NoStop}%
\bibitem [{\citenamefont {Witten}(1978{\natexlab{a}})}]{Witten:1978qu}%
  \BibitemOpen
  \bibfield  {author} {\bibinfo {author} {\bibfnamefont {E.}~\bibnamefont
  {Witten}},\ }\href {\doibase 10.1016/0550-3213(78)90416-9} {\bibfield
  {journal} {\bibinfo  {journal} {Nucl. Phys. B}\ }\textbf {\bibinfo {volume}
  {145}},\ \bibinfo {pages} {110} (\bibinfo {year}
  {1978}{\natexlab{a}})}\BibitemShut {NoStop}%
\bibitem [{\citenamefont {Berezinskii}(1971)}]{berezinskii1971destruction}%
  \BibitemOpen
  \bibfield  {author} {\bibinfo {author} {\bibfnamefont {V.~L.}\ \bibnamefont
  {Berezinskii}},\ }\href@noop {} {\bibfield  {journal} {\bibinfo  {journal}
  {Sov. Phys. JETP}\ }\textbf {\bibinfo {volume} {32}},\ \bibinfo {pages} {493}
  (\bibinfo {year} {1971})}\BibitemShut {NoStop}%
\bibitem [{\citenamefont {Kosterlitz}\ and\ \citenamefont
  {Thouless}(1973)}]{Kosterlitz:1973xp}%
  \BibitemOpen
  \bibfield  {author} {\bibinfo {author} {\bibfnamefont {J.~M.}\ \bibnamefont
  {Kosterlitz}}\ and\ \bibinfo {author} {\bibfnamefont {D.~J.}\ \bibnamefont
  {Thouless}},\ }\href {\doibase 10.1088/0022-3719/6/7/010} {\bibfield
  {journal} {\bibinfo  {journal} {J. Phys. C}\ }\textbf {\bibinfo {volume}
  {6}},\ \bibinfo {pages} {1181} (\bibinfo {year} {1973})}\BibitemShut
  {NoStop}%
\bibitem [{\citenamefont {Witten}(1978{\natexlab{b}})}]{Witten:1977xv}%
  \BibitemOpen
  \bibfield  {author} {\bibinfo {author} {\bibfnamefont {E.}~\bibnamefont
  {Witten}},\ }\href {\doibase 10.1016/0550-3213(78)90204-3} {\bibfield
  {journal} {\bibinfo  {journal} {Nucl. Phys. B}\ }\textbf {\bibinfo {volume}
  {142}},\ \bibinfo {pages} {285} (\bibinfo {year}
  {1978}{\natexlab{b}})}\BibitemShut {NoStop}%
\bibitem [{\citenamefont {Jackiw}\ and\ \citenamefont
  {Rebbi}(1976)}]{Jackiw:1975fn}%
  \BibitemOpen
  \bibfield  {author} {\bibinfo {author} {\bibfnamefont {R.}~\bibnamefont
  {Jackiw}}\ and\ \bibinfo {author} {\bibfnamefont {C.}~\bibnamefont {Rebbi}},\
  }\href {\doibase 10.1103/PhysRevD.13.3398} {\bibfield  {journal} {\bibinfo
  {journal} {Phys. Rev. D}\ }\textbf {\bibinfo {volume} {13}},\ \bibinfo
  {pages} {3398} (\bibinfo {year} {1976})}\BibitemShut {NoStop}%
\bibitem [{\citenamefont {Forgacs}\ \emph {et~al.}(1991)\citenamefont
  {Forgacs}, \citenamefont {Niedermayer},\ and\ \citenamefont
  {Weisz}}]{Forgacs:1991rs}%
  \BibitemOpen
  \bibfield  {author} {\bibinfo {author} {\bibfnamefont {P.}~\bibnamefont
  {Forgacs}}, \bibinfo {author} {\bibfnamefont {F.}~\bibnamefont
  {Niedermayer}}, \ and\ \bibinfo {author} {\bibfnamefont {P.}~\bibnamefont
  {Weisz}},\ }\href {\doibase 10.1016/0550-3213(91)90044-X} {\bibfield
  {journal} {\bibinfo  {journal} {Nucl. Phys. B}\ }\textbf {\bibinfo {volume}
  {367}},\ \bibinfo {pages} {123} (\bibinfo {year} {1991})}\BibitemShut
  {NoStop}%
\bibitem [{\citenamefont {Chodos}\ and\ \citenamefont
  {Minakata}(1997)}]{Chodos:1996pp}%
  \BibitemOpen
  \bibfield  {author} {\bibinfo {author} {\bibfnamefont {A.}~\bibnamefont
  {Chodos}}\ and\ \bibinfo {author} {\bibfnamefont {H.}~\bibnamefont
  {Minakata}},\ }\href {\doibase 10.1016/S0550-3213(97)00094-1} {\bibfield
  {journal} {\bibinfo  {journal} {Nucl. Phys. B}\ }\textbf {\bibinfo {volume}
  {490}},\ \bibinfo {pages} {687} (\bibinfo {year} {1997})},\ \Eprint
  {http://arxiv.org/abs/hep-th/9610150} {arXiv:hep-th/9610150} \BibitemShut
  {NoStop}%
\bibitem [{\citenamefont {Mari\~no}\ and\ \citenamefont
  {Reis}(2020)}]{Marino:2019eym}%
  \BibitemOpen
  \bibfield  {author} {\bibinfo {author} {\bibfnamefont {M.}~\bibnamefont
  {Mari\~no}}\ and\ \bibinfo {author} {\bibfnamefont {T.}~\bibnamefont
  {Reis}},\ }\href {\doibase 10.1007/JHEP04(2020)160} {\bibfield  {journal}
  {\bibinfo  {journal} {JHEP}\ }\textbf {\bibinfo {volume} {04}},\ \bibinfo
  {pages} {160} (\bibinfo {year} {2020})},\ \Eprint
  {http://arxiv.org/abs/1909.12134} {arXiv:1909.12134 [hep-th]} \BibitemShut
  {NoStop}%
\bibitem [{\citenamefont {Di~Pietro}\ \emph {et~al.}(2021)\citenamefont
  {Di~Pietro}, \citenamefont {Mari\~no}, \citenamefont {Sberveglieri},\ and\
  \citenamefont {Serone}}]{DiPietro:2021yxb}%
  \BibitemOpen
  \bibfield  {author} {\bibinfo {author} {\bibfnamefont {L.}~\bibnamefont
  {Di~Pietro}}, \bibinfo {author} {\bibfnamefont {M.}~\bibnamefont {Mari\~no}},
  \bibinfo {author} {\bibfnamefont {G.}~\bibnamefont {Sberveglieri}}, \ and\
  \bibinfo {author} {\bibfnamefont {M.}~\bibnamefont {Serone}},\ }\href
  {\doibase 10.1007/JHEP10(2021)166} {\bibfield  {journal} {\bibinfo  {journal}
  {JHEP}\ }\textbf {\bibinfo {volume} {10}},\ \bibinfo {pages} {166} (\bibinfo
  {year} {2021})},\ \Eprint {http://arxiv.org/abs/2108.02647} {arXiv:2108.02647
  [hep-th]} \BibitemShut {NoStop}%
\bibitem [{\citenamefont {Fateev}\ \emph
  {et~al.}(1994{\natexlab{a}})\citenamefont {Fateev}, \citenamefont
  {Wiegmann},\ and\ \citenamefont {Kazakov}}]{Fateev:1994dp}%
  \BibitemOpen
  \bibfield  {author} {\bibinfo {author} {\bibfnamefont {V.~A.}\ \bibnamefont
  {Fateev}}, \bibinfo {author} {\bibfnamefont {P.~B.}\ \bibnamefont
  {Wiegmann}}, \ and\ \bibinfo {author} {\bibfnamefont {V.~A.}\ \bibnamefont
  {Kazakov}},\ }\href {\doibase 10.1103/PhysRevLett.73.1750} {\bibfield
  {journal} {\bibinfo  {journal} {Phys. Rev. Lett.}\ }\textbf {\bibinfo
  {volume} {73}},\ \bibinfo {pages} {1750} (\bibinfo {year}
  {1994}{\natexlab{a}})}\BibitemShut {NoStop}%
\bibitem [{\citenamefont {Fateev}\ \emph
  {et~al.}(1994{\natexlab{b}})\citenamefont {Fateev}, \citenamefont {Kazakov},\
  and\ \citenamefont {Wiegmann}}]{Fateev:1994ai}%
  \BibitemOpen
  \bibfield  {author} {\bibinfo {author} {\bibfnamefont {V.}~\bibnamefont
  {Fateev}}, \bibinfo {author} {\bibfnamefont {V.}~\bibnamefont {Kazakov}}, \
  and\ \bibinfo {author} {\bibfnamefont {P.}~\bibnamefont {Wiegmann}},\ }\href
  {\doibase 10.1016/0550-3213(94)90405-7} {\bibfield  {journal} {\bibinfo
  {journal} {Nucl. Phys. B}\ }\textbf {\bibinfo {volume} {424}},\ \bibinfo
  {pages} {505} (\bibinfo {year} {1994}{\natexlab{b}})},\ \Eprint
  {http://arxiv.org/abs/hep-th/9403099} {arXiv:hep-th/9403099} \BibitemShut
  {NoStop}%
\bibitem [{\citenamefont {Kazakov}\ \emph {et~al.}(2024)\citenamefont
  {Kazakov}, \citenamefont {Sobko},\ and\ \citenamefont
  {Zarembo}}]{Kazakov:2023imu}%
  \BibitemOpen
  \bibfield  {author} {\bibinfo {author} {\bibfnamefont {V.}~\bibnamefont
  {Kazakov}}, \bibinfo {author} {\bibfnamefont {E.}~\bibnamefont {Sobko}}, \
  and\ \bibinfo {author} {\bibfnamefont {K.}~\bibnamefont {Zarembo}},\ }\href
  {\doibase 10.1103/PhysRevLett.132.141602} {\bibfield  {journal} {\bibinfo
  {journal} {Phys. Rev. Lett.}\ }\textbf {\bibinfo {volume} {132}},\ \bibinfo
  {pages} {141602} (\bibinfo {year} {2024})},\ \Eprint
  {http://arxiv.org/abs/2312.04801} {arXiv:2312.04801 [hep-th]} \BibitemShut
  {NoStop}%
\bibitem [{\citenamefont {Marino}\ \emph {et~al.}(2023)\citenamefont {Marino},
  \citenamefont {Miravitllas},\ and\ \citenamefont {Reis}}]{Marino:2023epd}%
  \BibitemOpen
  \bibfield  {author} {\bibinfo {author} {\bibfnamefont {M.}~\bibnamefont
  {Marino}}, \bibinfo {author} {\bibfnamefont {R.}~\bibnamefont {Miravitllas}},
  \ and\ \bibinfo {author} {\bibfnamefont {T.}~\bibnamefont {Reis}},\
  }\href@noop {} {\  (\bibinfo {year} {2023})},\ \Eprint
  {http://arxiv.org/abs/2302.08363} {arXiv:2302.08363 [hep-th]} \BibitemShut
  {NoStop}%
\bibitem [{\citenamefont {Gakhov}(1990)}]{Gakhov}%
  \BibitemOpen
  \bibfield  {author} {\bibinfo {author} {\bibfnamefont {F.}~\bibnamefont
  {Gakhov}},\ }\href@noop {} {\emph {\bibinfo {title} {{Boundary value
  problems}}}}\ (\bibinfo  {publisher} {Dover Publications},\ \bibinfo {year}
  {1990})\BibitemShut {NoStop}%
\bibitem [{\citenamefont {Gradshteyn}\ and\ \citenamefont
  {Ryzhik}(2014)}]{gradshteyn2014table}%
  \BibitemOpen
  \bibfield  {author} {\bibinfo {author} {\bibfnamefont {I.}~\bibnamefont
  {Gradshteyn}}\ and\ \bibinfo {author} {\bibfnamefont {I.}~\bibnamefont
  {Ryzhik}},\ }\href {https://books.google.dk/books?id=F7jiBQAAQBAJ} {\emph
  {\bibinfo {title} {Table of Integrals, Series, and Products}}}\ (\bibinfo
  {publisher} {Elsevier Science},\ \bibinfo {year} {2014})\BibitemShut
  {NoStop}%
\end{thebibliography}
\end{document}